# CME Acceleration as a Probe of the Coronal Magnetic Field

James Paul Mason[1], Phillip C. Chamberlin[1], Thomas N. Woods[1], Andrew Jones[1], Astrid M. Veronig[2],
Karin Dissauer[3], Michael Kirk[4], SunCET Team

[1]University of Colorado at Boulder, [2]University of Graz, [3]Northwest Research Associates, [4]NASA GSFC

## The Current State in 2020

Our community has an established understanding that the coronal magnetic field is the proximate cause of solar eruptive events. We have numerous instruments that observe enhanced plasma emission where the magnetic field strengthens into loops, e.g., EUV and X-ray imagers; although brightness drops rapidly with radial distance due to the emission measure's strong dependence on density, meaning most observatories have optimized for a smaller field of view while gaining a higher spatial resolution of the disk. We also have numerous models describing the coronal magnetic field to varying levels of fidelity, e.g., PFSS, NLFF, and MHD; although the lack of observations especially in the middle corona (~1.5-5 Rs) means a lack of strong constraints for magnetic field topology and dynamics (Figure 1).

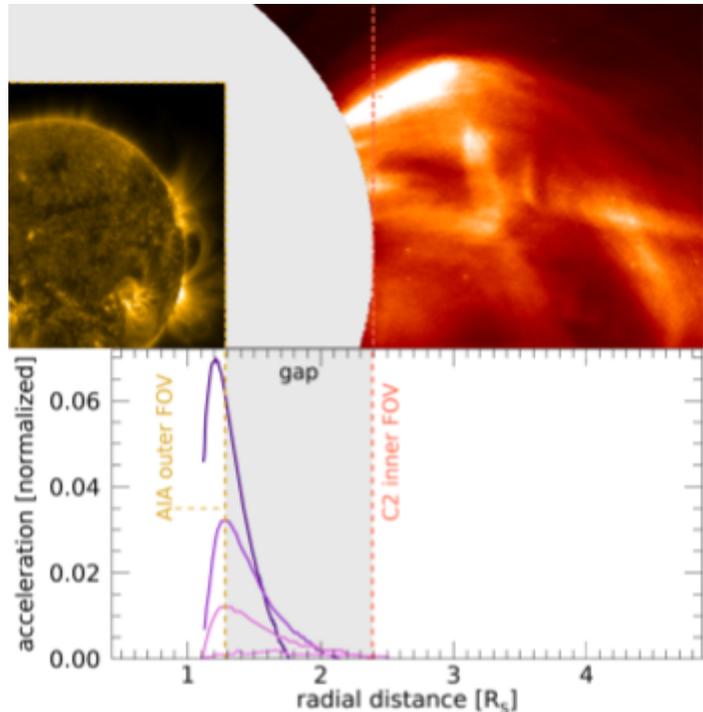

Figure 1: Top: Composite of SDO/AIA 171 Å image and SOHO/LASCO/C2 white-light coronagraph image. Bottom: Modeled acceleration profiles of torus-instability CMEs, adapted from Török and Kliem (2007). The different curves result from different background magnetic field assumptions. Most of the acceleration occurs in the under-observed and thus under-constrained region of the middle corona.

There are at least 26 *review* papers on the topic of CME trigger and propagation mechanisms over the last two decades (Green *et al.* 2018, and references therein) -- a testament to the community's sustained, intense interest and the fact that there remain many open questions. The standard model of a CME eruption puts an upward force in opposition by a constraining force provided by the surrounding magnetic field. Perhaps the simplest model of this defines a 1D, horizontal background magnetic field that declines in strength with height, characterized by the "decay index" of the torus instability model (Bateman, 1978; Kliem & Török, 2006; Török & Kliem, 2007). As Figure 1 shows, even this simple model can have very different CME acceleration profiles by changing the value of the decay index. Adding a simple upward velocity perturbation with tunable duration is sufficient to change the fundamental shape of the acceleration profiles from single peak to double peak (Schrijver *et al.* 2008). Here too, these profiles distinguish themselves in the middle corona. Several more sophisticated models also focus on removing overlying field constraints, including breakout, slip-running reconnection, and reconnection in the current sheet. The helical kink instability focuses on the magnetic field topology of the flux rope, which simulations show also has acceleration profiles varying through the middle corona (e.g., Fan 2016). The CME acceleration profile is also dependent on the 3D structure of the erupting material and the surrounding field, potential drainage of dense plasma, and continued magnetic reconnection freeing more energy to drive the CME. The other aspect of acceleration is direction: CMEs can be deflected away from a "pure" radial propagation by as much as ~30º, which is again primarily determined by the surrounding coronal magnetic field strength in the middle corona (Figure 2).

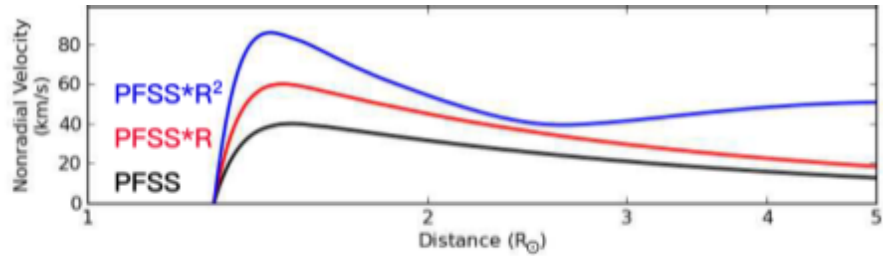

Figure 2: ForeCAT simulations of a CME propagating through background magnetic fields (PFSS) of various strengths (Kay 2016). R is radial distance. CMEs experience greater non-radial velocity in middle corona environments with stronger magnetic fields.

Given observational constraints, the models must get the 3D magnetic field configuration and dynamics in the low and middle corona correct in order for the modeled kinematic profiles to match the observed ones. Thus, CME kinematic profiles and the coronal magnetic field are tightly coupled.

**The Desired State by 2050**

**By 2050, we expect that CME models will accurately describe, and ideally predict, observed solar eruptions and the propagation of the CMEs through the corona.** Models are the manifestation of our theoretical understanding of natural phenomena; therefore when this milestone is reached we can be confident in our understanding of the coronal magnetic field. We anticipate that this understanding can be distilled into relatively simple models that can describe most cases sufficiently well for space weather forecasting purposes, while the more accurate models will be reserved for detailed scientific study or specialized prediction purposes. This is the case for Newtonian gravity and general relativity and is largely the state today for CMEs and the magnetic field; the difference will be in the accuracy of the models and the depth of our understanding.

**The Journey**

In order to reach the desired state, we must address the known unknowns and be prepared for the unknown unknowns. The below lists are not intended to be comprehensive, but hopefully offer some useful ideas.

Addressing the known unknowns:
- Expand observational regimes.
  - Use direct coronal magnetic field observations to constrain models. Such observations are presently sparse. There will soon be more from DKIST, but still far from synoptic.
  - Observe the sun as it is: a 3D object requiring multiple observational angles and use those observations to constrain models. See for example the white paper "The Science Case for the 4π Perspective: Studying the Evolution & Propagation of the Solar Wind and Solar Transients" led by Angelos Vourlidas and the white papers discussing polar observations led by Jeff Newmark and Sarah Gibson.
  - Observe the middle corona to constrain models. The case for this has been made briefly here and is expanded upon in Mason 2020.
- Improve upon the existing CME models
  - More complete treatment of fundamental plasma and MHD processes, rather than focuses on just a few key ones.
  - Incorporate new constraints provided by new observations from PSP, SO, and any other expanded observational regimes.
  - Link magnetic field lines from the photosphere to the source surface
  - Achieve consistent results between multiple runs in the solar max regime

Preparing for the unknown unknowns:
- Expand the community. There are many observable sun-like stars. There is likely a lot to be gained by collaborating with astronomers on this topic. See the white paper "Solar Analogs as a Tool to Understand the Sun" led by Allison Youngblood.
- Make powerful tools, such as machine learning, easy to use and big data easy to access and manipulate even for non-experts in heliophysics and/or in computer science. See white papers "Exploring the Critical Coronal Transition Region: The Key to Uncovering the Genesis of the Solar Wind and Solar Eruptions" led by Angelos Vourlidas and "Science Platforms for Heliophysics" led by Will Barnes.